\documentclass[prl,aps,amssymb,amsfonts,amsmath,showpacs,twocolumn,superscriptaddress]{revtex4}
\usepackage{graphicx}

\begin{document}
\newcommand{\ethaffil}{Laboratory of Physical Chemistry, Swiss Federal Institute of Technology (ETH), 8093 Zurich, Switzerland}
\newcommand{\zuseaffil}{Zuse Institute Berlin, Takustrasse 7, 14195 Berlin, Germany.}
\newcommand{\potsdamaffil}{Institute of Physics, University of Potsdam, 14469 Potsdam,
Germany.}

\title{Optical microscopy via spectral modifications of a nano-antenna}

\author{T.~Kalkbrenner}
\altaffiliation[Present address: ]{ FOM-Institute for Atomic and
Molecular Physics, 1098 SJ Amsterdam, The Netherlands}
\affiliation{\ethaffil}
\author{U.~H\aa kanson}
\affiliation{\ethaffil}
\author{A.~Sch\"{a}dle}
\affiliation{\zuseaffil}
\author{S.~Burger}
\affiliation{\zuseaffil}
\author{C.~Henkel}
\affiliation{\potsdamaffil}
\author{V.~Sandoghdar}\email{vahid.sandoghdar@ethz.ch}
\affiliation{\ethaffil}

\pacs{07.79.Fc, 42.50.Lc, 42.30.-d, 78.67.Bf}

\begin{abstract}
The existing optical microscopes form an image by collecting
photons emitted from an object. Here we report on the experimental
realization of microscopy without the need for direct optical
communication with the sample. To achieve this, we have scanned a
single gold nanoparticle acting as a nano-antenna in the near
field of a sample and have studied the modification of its
\textit{intrinsic }radiative properties by monitoring its plasmon
spectrum.
\end{abstract}

\maketitle

\newpage

Over the years, several clever techniques such as dark-field,
phase contrast, fluorescence, differential interference contrast,
confocal, and scanning near-field microscopies have provided
powerful ways of performing optical imaging. In all these methods,
as in any other visual process, one "sees" an object when photons
originating from it reach the detector. The details of the imaging
mechanism depend sensitively on the intensity, phase and
polarization of light both in the illumination and collection
channels. The thought of recording optical images \textit{without}
receiving photons from the object, therefore, seems to be a
contradiction in terms. In this Letter we show that this is indeed
possible if one monitors the intrinsic spectral properties of a
nanoscopic antenna scanned close to the sample.

When an oscillating dipole is placed in confined geometries its
radiative properties, such as eigenfrequency and linewidth, are
modified \cite{Barnes:98,Berman}. In an intuitive picture these
modifications are due to the interaction of the oscillating dipole
with its image dipoles whereby the boundary materials and their
distances to the oscillator determine the strength and phase of
this interaction. In an alternative point of view the radiative
changes are due to the modification of the density of photon
states available for emission. In the context of recent
developments in nano-optics, theoretical investigations have
extended these concepts to subwavelength geometries and have shown that the linewidth \cite%
{Novotny:96,Rahmani:97,Henkel:98,Quinten:98,Parent:99} and the
transition frequency \cite{Henkel:98,Quinten:98} of a dipole also
respond sensitively to the optical contrast of its nanoscopic
environment. Hence, it has been proposed that the spectral
modifications of a nano-antenna could serve
as the signal for a novel mode of Scanning Near-field Optical Microscopy (SNOM)~\cite%
{Rahmani:97,Henkel:98,Parent:99}. Here we report on the first
realization of this idea, using a single gold nanoparticle as a
subwavelength antenna.

It is well known that the collective excitation of free electrons
in metallic particles can give rise to plasmon resonances,
depending on the size, shape, and index of refraction of the
particle as well as the optical constants of its surrounding
\cite{Kreibig}. For a spherical particle placed in a homogenous
medium, the plasmon spectrum can be calculated using Mie theory
based on a multipole expansion. For gold particles of diameter 100
nm or smaller the optical response is by large due to the dipolar
mode of the plasmon excitation. Thus, a gold nanoparticle can be
considered as a dipolar antenna with a well-defined resonance
frequency and linewidth. Having a large polarizability, single
gold nanoparticles can be optically detected even down to a size
of about 5~nm \cite{Lindfors:04}. Furthermore, the large quantum
yield of a gold nanoparticle~\cite{buchler:05} makes it a suitable
candidate for studying the radiative properties of its
environment. Finally, compared to the commonly used systems such
as single dye molecules or semiconductor nanocrystals, metallic
nanoparticles have the decisive advantage that they are
indefinitely photostable.

\begin{figure}[b!] \centering
\includegraphics[width=8.5 cm]{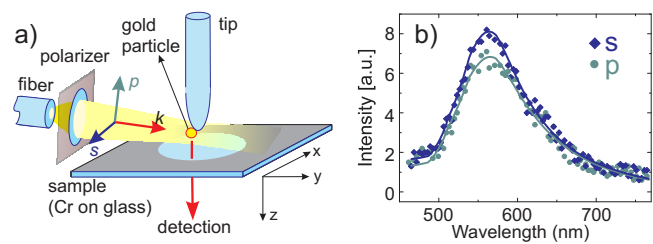}
\caption{a) Schematics of the experimental arrangement. b) Plasmon
spectra of the gold nanoparticle attached to the glass fiber tip
in the absence of the sample. The polarization of the incident
light is indicated by s and p defined in a).} \label{setup}
\end{figure}

In our laboratory we have developed a procedure for mounting a
single gold nanoparticle to the end of an uncoated glass fiber
tip~\cite{Kalkbrenner:01}. We have also reported on the
realization of apertureless SNOM by detecting the intensity of the
light scattered from the gold nanoparticle at a fixed
wavelength~\cite{Kalkbrenner:01}. In the current experiment we use
the same kind of probe (particle diameter of 100~nm), but we now
record its plasmon spectrum at every scan pixel. Figure 1a shows
the principal elements of the experiment. White light from a xenon
lamp is brought to the experimental stage via a multimode fiber
with a core diameter of $100~\mu m$, passed through a polarizer
and weakly focused onto the tip end by using a lens system. The
scattered light is collected by a microscope objective, directed
through a pinhole, coupled into a multimode fiber, and sent to a
spectrometer equipped with a cooled CCD camera. Figure 1b displays
scattering spectra of a gold nanoparticle at the end of the tip
after it was picked up but away from the substrate. The diamonds
and circles show the recorded spectra for \emph{s} and \emph{p}
polarized incident light, respectively. The fact that the
\textit{s} and \textit{p} spectra are nearly identical lets us
conclude that the cross-section of the particle facing the
incident beam is to a very good approximation
circular~\cite{Kalkbrenner:04}.

A suitable sample for our studies should have a minimal
topography, large optical contrast with sharp edges and be
sufficiently transparent to allow detection in transmission. To
fabricate such a sample, first we spin coated latex spheres of
diameter $2~\mu m$ on a microscope cover slip at a very low
coverage. Next we evaporated about $8~nm$ of chromium and removed
the latex beads. The resulting sample consists of $2~\mu m$ round
openings in a semi-transparent Cr film on a cover glass with very
sharp edges that rise within less than
10~nm~\cite{Kalkbrenner:01}. This sample was placed on a 3D
piezoelectric scanner, mounted on an inverted microscope. A
home-made SNOM stage was used to position the tip and to stabilize
its distance from the sample to about 5-10~nm via shear-force
feedback~\cite{barenz:96}. Figure 2a shows the shear-force
topography image recorded when the sample was scanned under the
gold nanoparticle probe, whereas the inset displays a cross
section along cut (ii) (see Fig.~2c). The elevated annulus (height
of about 30~nm) in the middle of the glass region is produced by
the collection of residual material at the junction between the
latex particles and the substrate in the spin casting and lift-off
processes.

During the shear-force scan shown in Fig.~2a the spectrometer
camera was triggered at each of the 1600 pixels, and the plasmon
spectrum of the gold particle was recorded. The symbols in Fig.~3a
show spectra from pixels $\alpha $ and $\beta $ marked in Fig.~2a.
Fig. 2b displays the total intensity integrated over the
wavelength range 450-750~nm for each pixel. As expected, the
signal is higher when the gold nanoparticle is on the glass part
of the sample~\cite{Kalkbrenner:01}, but the intensity
distribution is somewhat smeared out at the edges because we have
not discriminated against unwanted far-field stray scattering.
Efficient separation of the near-field signal from the stray
background is also a central issue in conventional apertureless
SNOM using extended
tips~\cite{Zenhausern:95,Inouye:94,gleyzes:95,Labardi:00,keilmann:04}.

\begin{figure}[t!]
\centering
\includegraphics[width=8.5 cm]{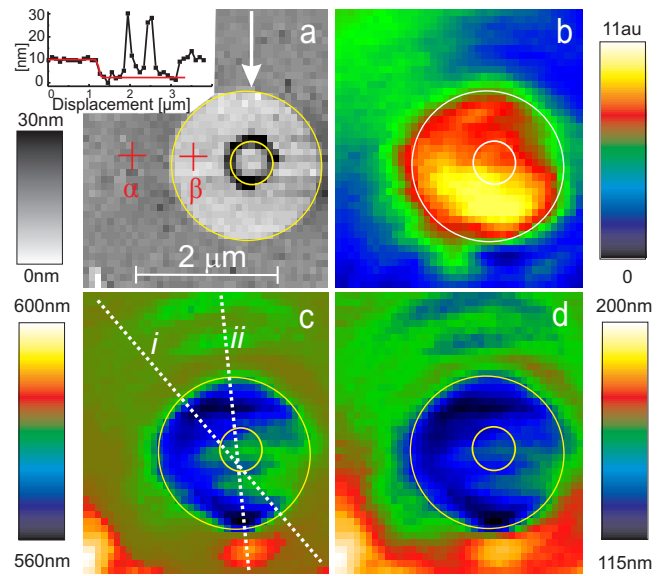}
\caption{a) Topography image of the sample ($40\times40$ pixels),
showing the round opening in a very thin chromium film. Inside
this opening a small round protrusion is revealed (see text for
details). The inset shows a cross section along cut (ii). The
arrow indicates the direction of illumination. b) The transmitted
intensity integrated under the spectrum recorded at each pixel. c)
and d) The central wavelength and the FWHM of the plasmon
resonance, respectively. The circles are traced as a guide to the
eye.}\label{data}
\end{figure}

Now we discuss the essential results of our work, namely the
modifications of the particle's plasmon spectrum. Comparison of
the spectra in Fig.~3a reveals a red shift and a broadening of
$\alpha $\ with respect to $ \beta $. While it is not possible to
state a simple analytic formula for the spectrum of the
nanoparticle in the near field of an inhomogeneous sample, we have
found that we can fit our experimental data using a
phenomenological model. In order to quantify the observed spectral
modifications, we adapted the formalism of Wokaun, Gordon and
Liao~\cite{Wokaun:82} based on the definition of an effective
polarizability for a metallic nanoparticle. These authors showed
that the total induced dipole moment $P(\omega )$ of a
nanoparticle with volume $V\ll \lambda ^{3}$ and dipolar
polarizability $\alpha (\omega )$ is given by
\begin{equation} P(\omega )=\alpha(\omega)\left[
E_{0}+i\eta \left( 2k^{3}/3\right) P(\omega )\right]
\label{dipole}
\end{equation}
where $E_{0}$ denotes the applied external electric field and
$k=2\pi/\lambda$. The first term in Eq.~(\ref{dipole}) considers
the dipole moment that is induced directly by the incident field.
The second term describes the so-called radiation damping effect
where the field emitted by the particle interacts with its own
dipole moment $P(\omega )$. In the original work of Wokaun and
coworkers $\eta$ was set to 1, but here we have introduced this
real parameter in a phenomenological manner to represent the part
of the particle's self field caused by reflections from the
surroundings. We note that since we are dealing with near-field
interactions, retardation effects could be neglected at this
point. Rearrangement of Eq.~(\ref{dipole}) leads to an expression
for the effective dipolar polarizability,
\begin{equation}
\alpha_{eff}(\omega )=\frac{P(\omega )}{E_{0}}=\frac{\alpha
(\omega )}{1-\eta [i2k^{3}\alpha (\omega )/3]}. \label{susc}
\end{equation}%

\begin{figure}[t!]
\centering
\includegraphics[width=8.75 cm]{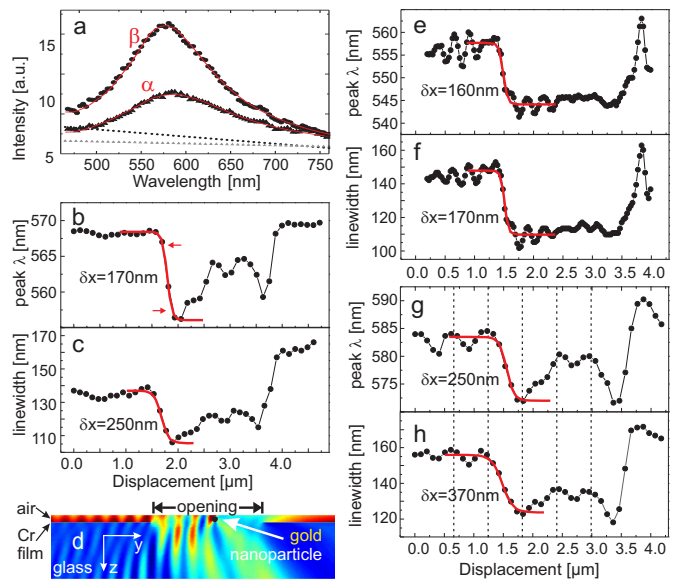}\\
\caption{a) Examples of the spectra recorded on the chromium and
glass regions at positions $\alpha$ and $\beta$ in
Fig.~\ref{data}. Peak wavelengths and linewidths from cross
sections \emph{i} and \emph{ii} in Fig.~\ref{data} are shown in
(b, c) and (g, h), respectively. d) An example snap shot of the
simulated distribution of intensity $\left\vert H_x \right\vert
^{2}$. Due to lack of space, we only show the central part of the
computation box. e) and f) Simulated peak wavelength and linewidth
of the plasmon spectra for a 2D system. The red curves show fits
by the edge function $tanh((y-y_0)/l)$ where $y_0$ and $l$ are
free parameters. In each case the parameter $\delta x$ represents
the width from the 90\% to 10\% value of this
function.}\label{analysis}
\end{figure}

The solid curves in Fig.~1b represent theoretical fits to the
particle plasmon resonance in the absence of a substrate; i.e.
$\eta$ was set to 1. Here the detected intensity was taken to be
proportional to the scattering cross section and therefore to
$k^{4}\left\vert \alpha_{eff} (\omega )\right\vert ^{2}$, using
the quasi-static polarizability
$\alpha(\omega)$~\cite{Kreibig,Kalkbrenner:04}. Following the same
procedure, but leaving $\eta $ as a variable parameter, we then
fitted the plasmon spectra of the gold particle in the immediate
vicinity of the sample. The solid curves in Fig.~3a show examples
of the fits obtained for spectra $\alpha $ and $\beta$. The dashed
lines display the linear functions that have been added to the
radiated intensity of the particle to account for the background
scattered light. The excellent signal-to-noise ratio and the
robust fit to the experimental spectra allow us to determine even
small changes in the peak positions and the full widths at half
maxima (FWHM) of the spectra. The results for all scan pixels are
displayed in Figs.~\ref{data}c and d. The circular glass opening
in the chromium film is clearly imaged in both cases.
Figures~\ref{analysis}b and c display cross sections of
Figs.~\ref{data}c and d along cut (\emph{i}). As expected, the
near-field coupling of the antenna to the lossy excitations in the
chromium film causes a red shift and a broadening of its resonance
when it is moved from the glass region to above the
Cr-film~\cite{chance:78}. These results embody the essence of this
novel near-field imaging mechanism where the information about the
local optical contrast of the sample is "encoded" in the spectrum
of the nano-antenna and not in the field, phase or polarization of
the received photons.

We now discuss some details of the images in Fig.~\ref{data}. To
facilitate this, we have overlayed circles as guides to the eye.
These were first matched to the circular features of the
topography image and then overlayed onto images~\ref{data}c and d,
allowing for a lateral offset due to a commonly observed
displacement between the position of the gold nanoparticle and the
lowest edge of the tip that traces the topography. We note that
images~\ref{data}c and d are much sharper than that in 2b where
the overall scattered intensity is plotted. Furthermore, it is
interesting that the central circular sedimentation does not
appear in Figs.~\ref{data}c and d, presumably because its
refractive index is not very different from that of glass. There
are, however, modulations of the resonance wavelength and
linewidth in the illumination direction, both on the Cr-film and
within the glass opening. We will discuss these below, but before
doing so, we present numerical calculations that provide
quantitative agreement with the magnitudes of the observed
spectral shift and broadening.

\enlargethispage*{0.5cm}

Consideration of experimental parameters such as the finite size
of the gold particle, its distance to the sample, the thinness of
the Cr-film (less than a skin depth) and the interface between
chromium and glass calls for rigorous numerical simulations. To do
this, we have used the finite-element method to solve the
time-harmonic Maxwell's equations in a two-dimensional setting for
the magnetic field component $H_x$ perpendicular to the
computational plane. We took the tabulated dielectric functions
for gold~\cite{JohnsonChristy:72} and chromium
\cite{JohnsonChristy:74} and implemented the Sommerfeld radiation
condition with a special perfectly matched layer adapted to the
structure of the experiment~\cite{Zschiedrich:05}.
Figure~\ref{analysis}c displays a snap shot of the intensity
distribution $\left\vert H_x \right\vert ^{2}$ at $\lambda=570$~nm
for a particular particle position. We determined the energy flux
scattered into the detection numerical aperture from the angular
spectrum representation of the field at the lower border of the
simulation box, which was $7.4~\mu m$ wide. By repeating the
calculations for a large number of wavelengths, we obtained
spectra that we fitted using the same procedure used for the
experimental spectra. Figures~\ref{analysis}e and f display the
central wavelength and the FWHM of the calculated spectra as a
function of particle position. For comparison, in
Figs.~\ref{analysis}g and h we also show the experimental data
along the illumination direction corresponding to cut (\emph{ii})
in Fig.~\ref{data}. In both the theoretical and experimental
curves we find about 30~nm broadening and 10~nm of red shift in
the plasmon resonance as the particle is moved from the glass to
the chromium section of the sample. The agreement between the
experimental and theoretical findings is quantitative even though
the calculations were only two-dimensional.

Next, we turn to the issue of modulations in Figs.~\ref{data}c and
d, which appear on both Cr and glass sections. The dashed lines in
cross sections~\ref{analysis}g and h indicate that these
modulations repeat at a period of about 580~nm, which is of the
order of the particle plasmon resonance wavelength
$\lambda_{res}$. The key issue is that although we have minimized
the collected stray light by illumination at grazing incidence and
by placing a pinhole in the detection path, the edges of the
circular opening can scatter a non-negligible amount of light onto
the detector. This background field interferes with the radiation
of the particle at the detector (note that as shown in
Fig.~\ref{analysis}a, the background is much smaller than the
particle radiation). As described in the supplementary
material~\footnote{See supplementary material for a
detailed description of this effect.}, scanning the sample results
in an apparent modulation of the peak wavelength and linewidth of
the plasmon spectrum at a period of about $\lambda_{res}$.
Interestingly, Figs.~\ref{analysis}e and f also display periodic
modulations but this time at a period of about $\lambda_{res}/2$.
Indeed, as indicated in Fig.~\ref{analysis}d, there are strong
standing waves that can be traced to the interference of the
particle backscattering and the illumination fields around the
sample. These fast modulations are, however, an artefact of 2D
considerations which overestimate backscattering as compared to 3D
geometries. Finally, we have checked that due to the large
absorption of Cr, the coupling of the particle emission to surface
plasmons~\cite{hecht:96} in the Cr film plays a negligible role in
all these processes.

As a last point, we remark that the spatial resolution of our
method depends on the size of the antenna. In our current
experiment, however, the observed edge sharpness of 170-370~nm in
Figs.~\ref{analysis} have been strongly influenced by the
above-mentioned modulations, preventing us from determining the
true resolution of the system. This source of artefact could be
fully eliminated if imaging were performed via the measurement of
the excited-state lifetime of a nanoscopic fluorescent emitter.

In conclusion, we have demonstrated that the intrinsic spectral
properties of a nano-antenna can be used to probe the near-field
optical contrast of a sample. The interesting aspect of this novel
experiment is that there is neither a need to illuminate the
sample, nor to receive light from it; instead one has to excite a
subwavelength radiating dipole and measure the changes in its
radiative characteristics. Although in the present work we have
used photons to "read out" the spectrum of the antenna, one could
imagine achieving this via electronic excitation of plasmon
resonances~\cite{Degiron:04} or by using electrical excitation of
luminescent material~\cite{Kuck:92,Baier:04}. Changes of polariton
or plasmon resonances in scanning probe microscopy have been also
addressed in recent experiments where a tip interacts with an
extended open geometry ~\cite{aizpurua:02,Taubner:04}. Here we
have realized an experiment where the modification of a
well-defined \emph{localized} plasmon resonance due to its
near-field interaction with the sample has been put into evidence.
Future experiments using very small gold
nanoparticles~\cite{Lindfors:04} or fluorescent emitters promise
to push the resolution of our method to the nanometer level.

\enlargethispage*{0.5cm}

We thank B. Buchler and S. K\"{u}hn for fruitful discussions. This
work was supported by the Swiss National Foundation (SNF), ETH
Zurich, Deutsche Forschungsgemeinschaft (DFG) priority program SPP
1113 and the DFG Research Center {\sc Matheon}.

\newpage
\noindent Supplementary material for \textit{"Optical
microscopy using the spectral modifications of a nano-antenna"} by
T. Kalkbrenner, U. H\aa kanson, A. Sch\"{a}dle, S. Burger, C.
Henkel, and V. Sandoghdar\\

We treat the issue of modulations observed both in peak position
and linewidth along the direction of illumination, both on the Cr
and glass parts of the sample (see Figs. 2c, 2d, and 3g and 3h of
the Letter). We believe that the modulation is caused by the
interference between the particle radiation and the stray light
scattered by the edge(s) of the chromium film. In our experiment
the illumination and therefore the light scattered by the edges
are white. The light scattered by the particle, on the other hand,
follows its plasmon spectrum. Figure 4a below shows the schematic
essence of the situation at hand, where we have replaced the
sample by a localized white scatterer (this is justified because
the pinhole in our detection path only selects light from a very
small region of the sample). Let us consider the gold particle to
be fixed. For simplicity, we can take its radiation spectrum to be
a Lorentzian of width $\gamma$, centered at frequency $\omega_0$
(see Fig. 4b). The electric field of the light scattered by the
particle can be written,
\begin{equation}
E_{part}(\omega)=A_{part}\frac{\gamma/2}{\gamma/2-i(\omega-\omega_0)}.
\label{lorentzian}
\end{equation}%
The electric field of the background light can be written as,
\begin{equation}
E_{bg}(\omega, z)=A_{bg}e^{i\omega z /c},\label{bg}
\end{equation}
where $z$ is defined as the path difference between the particle
emission and that of the background light (see Fig. 4a). Here we
have taken the paths from the gold particle and the white
scatterer to be about equal, which is a very good assumption for a
far detector. Note that $z\approx y$ for grazing incidence. The
total intensity at the detector is thus given by
\begin{equation}
I_{tot}(\omega, z)=|E_{part}( \omega ) + E_{bg}( \omega,
z)|^2.\label{bg}
\end{equation}

\noindent To approximate our experimental conditions, we take
$\gamma=0.2 \omega_0$ and consider the intensity of the background
light to be about 10\% of the maximum particle radiation. Now we
let $z$ vary (i.e. we scan the sample). Each wavelength component
of the background field and the particle field interfere
differently, leading to a slight modification of the line shape.
Fig.~4c shows examples of the resulting spectrum for four distinct
path differences. Fig.~4d shows the same information in grey
scales over a displacement of about $3\lambda_0$. Note that
although the resulting spectra are no longer truly Lorentzian, one
can still attribute a line center and FWHM to them. 
\clearpage 
Figures 4e and
f show the FWHMs and peak frequencies of these spectra, which turn
out to oscillate by a small amount. The oscillations are not
periodic, but for small displacements one can recognize a period
of about $\lambda_0=\omega_0 /2\pi c$. This picture is in very
good agreement with the observation of the small modulations in
our experimental data.

\begin{figure}[h!]
\centering{
\includegraphics{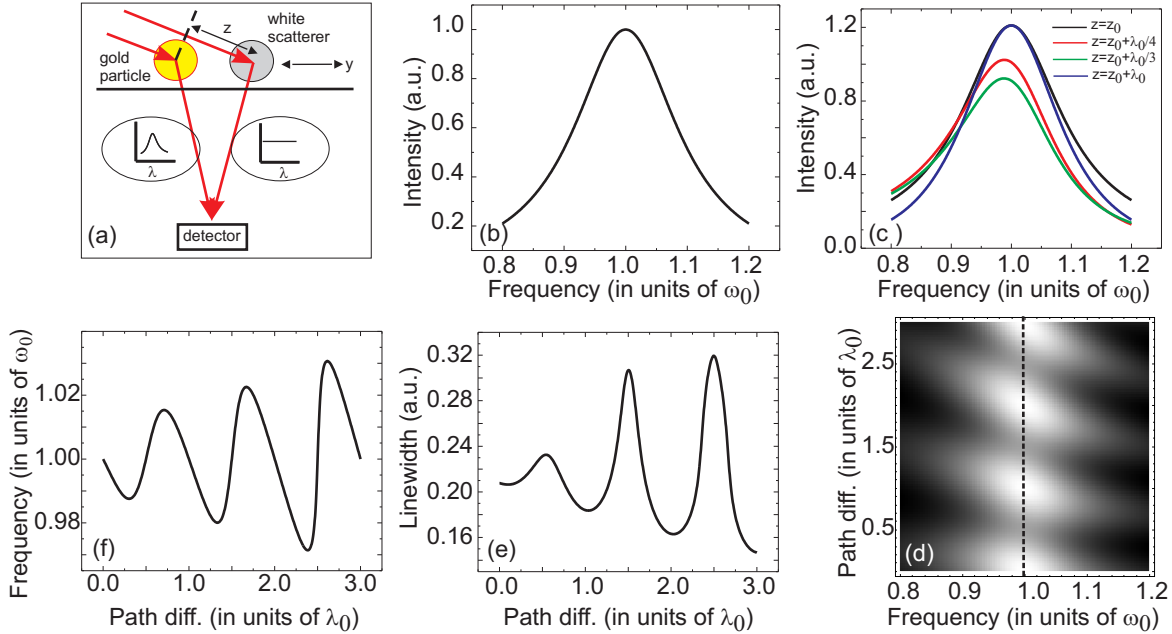}}
\caption{(a) Schematics of the situation. (b) A broad Lorentzian
spectrum. (c) Four examples of the resulting spectra. $z_{0}$
denotes the zero path difference. (d) Contour plot of the total
intensity as a function of the frequency and path difference. The
linewidth (e) and peak position (f) along the cross section in
(d).}
 \label{fig1}
\end{figure}

\end{document}